\documentclass[11pt]{article}

\usepackage{graphicx}
\usepackage{amsmath}
\usepackage{amsfonts}
\usepackage{amssymb}

\numberwithin{equation}{section}
\usepackage{amsmath}

\usepackage{diagrams}
\newarrow{XK}{|}{-}{-}{-}{triangle}
\newarrow{Eq}{=}{=}{=}{=}{=}

\advance \textheight by \topskip

\begin{document}

\markboth{L. J. Boya and R. Campoamor-Stursberg} {COMPOSITION
ALGEBRAS AND THE TWO FACES OF $G_{2}$}

\title{Composition algebras and the two faces of $G_{2}$}

\author{Luis J. BOYA\\
Departamento de F\'{\i}sica Te\'orica,\\Universidad de
Zaragoza,\\
E-50009 Zaragoza, Spain.\\
luisjo@unizar.es\\
\\
Rutwig Campoamor-Stursberg\\
Instituto de Matem\'atica Interdisciplinar and\\
Universidad Complutense de Madrid\\
3, plaza de Ciencias, E-28040 Madrid, Spain\\ rutwig@pdi.ucm.es}
 
\maketitle

\begin{abstract}
We consider composition and division algebras over the real
numbers: We note two r\^{o}les for the group $G_{2}$: as
automorphism group of the octonions and as the isotropy group of a
generic $3$-form in $7$ dimensions. We show why they are
equivalent, by means of a regular metric. We express in some
diagrams the relation between some pertinent groups, most of them
related to the octonions. Some applications to physics are also
discussed.
\end{abstract}


\section{Introduction}

The first exceptional Lie group $G_2$ (dimension 14, rank 2) was
first discovered, as its complex Lie algebra, in 1887 by W.
Killing, in his (redundant) list of all complex simple Lie
algebras; the list was fixed by \'E. Cartan in his Paris Thesis
(1894). $G_2$ is the only Lie group which has a Dynkin diagram
with a triple bond, $\bullet\equiv\circ$: the two smallest
representations ($\neq$ Id) are of dimensions 7 and 14. In physics
the group $G_2$ appears in several context, as in $7$-dim
compactifications from M-theory.

On the other hand, the division algebra of the octonions was first
written by J. Graves in Christmas, 1843 (letter to W. Hamilton,
who soon realized they were non-associative); however, the
eight-squares sum, related also to octonions (as the four- squares
sum (Euler, 1738) was related to the quaternions), had been found
earlier by C. Degen (1818); in 1845 A. Cayley also (re)discovered
the octonions in relation to his work on hyperelliptic functions;
so the name Cayley numbers, octaves and octonians are also used in
the literature for the octonions.

\'E. Cartan was the first to consider the group $G_2$ (in its real
compact form) as the automorphism group of the octonion algebra
(1914; it is quoted even in an earlier article of his in 1908); he
also showed there exists another, noncompact real form, that today
we know it is the automorphism group of the split octonions. But
already in 1900, F. Engel (who wrote with S. Lie the monumental
3-volume book on Transformation Groups \cite{Lie}) established
(the complex form of) $G_2$ as the isotropy group of a generic
$3$-form in $7$-dimensional complex space; his disciple W. Reichel
worked out the details of this description of $G_2$, including the
two real forms, in 1907. In modern times (ca. 1984) this $3$-form
has been much used by Bryant \cite{Bry} in relation to manifolds
of $G_2$ holonomy. All this historical record is narrated in
detail in \cite{Baez} and \cite{Agric}, see also \cite{Rosen,
Rosen2}.

This paper is devoted to the relation between composition and
division algebras and the two approaches for $G_2$; indeed, the
two conceptions ($G_2$ = Aut (Octonions) and $G_2$ = isotropy
group of a generic $3$-form in $7$ dimensions) have been already
noticed as equivalent by \cite{Agric,Baez} and others, but it
seems to us rather obscure the stated relations (one exception is
\cite{Hit}). Specifically, we shall construct the octonions
starting with the regular $3$-form $\phi$, wherefrom the dual role
of $G_2$ will be evident: the $3$-form determines a regular metric
$\beta$, which converts the form $\phi$, as $(0,3)$-tensor, in a
$(1,2)$-tensor: that is, an algebra  $\gamma =\beta(\phi)$, which
turns out to be, of course, the octonion algebra: the group $G_2$
becomes automatically the automorphism group of that algebra. In
the work of Bonan \cite{Bon}, a reverse result seems to be stated:
starting from Cartan's description of compact $G_2$ as the
automorphism group of the octonions, he saw how to derive from its
multiplicative law an invariant $3$-form. Before this we shall
elaborate on composition and division algebras and then also
extend some remarks of \cite{Agric} on the relations of Spin(8),
Spin(7) (both have a $8$-dim real representation) and other
smaller groups, including $G_2$, to the above work. Several
physical applications will be also briefly mentioned, including
some r\^{o}le for $SU(3)$ connected with octonions.

But first we need to recall some important fact about reals
$\mathbb{R}$, complex $\mathbb{C}$, quaternions $\mathbb{H}$ and
octonions $\mathbb{O}$, and their split forms (for the last
three).

\section{Complex and quaternions as composition algebras}

If $\mathbb{F}$ is a field of numbers, an algebra $A =
A(\mathbb{F})$ is a vector space over $\mathbb{F}$ with a linear
map: $A\times A\rightarrow A$ distributive with respect to
addition. We shall not assume commutativity  nor associativity in
the algebra. Typical examples are the complex numbers $\mathbb{C}$
over the reals $\mathbb{R}$, or the generic linear maps $End(V)$
for $V$ an $n$-dim vector space over $\mathbb{F}$.

An algebra is called composition algebra, if there is a bilinear
regular symmetric form  $\beta: A\times A\rightarrow \mathbb{F}$
whose associated quadratic form $Q(a):=\beta(a, a)$ satisfies
$Q(ab)=Q(a)Q(b)$; we shall usually take $\mathbb{F}= \mathbb{R}$
or $\mathbb{C}$. It is called division algebra if $ab=0$ implies
either $a=0$ or $b=0$; it is called normed algebra if it has a
norm as a vector space $N: A\rightarrow \mathbb{R}$ verifying
$N(a\neq 0) > 0$ and $N(ab)= N(a)N(b)$; if $A$ is division
algebra, then $Q=Q(A)$ is definite, and $N$ is $+\sqrt{Q}$: only
division algebras can be normed, although sometimes the norm is
defined as $N =+\sqrt{\left|Q\right|}$.

Let us briefly recall the composition algebras over the reals,
$\mathbb{F} = \mathbb{R}$. An algebra on $\mathbb{R}^2$ is
determined once $J^2$ is known, where $J:= \{0,1\}$ is a second
unit vector, besides $1 =\left\{1,0\right\}$. There are three
cases \cite{Rosen}: if $J^2=0$, we have the degenerate complex
numbers $\mathbb{C}^0$; we shall not consider them any further. If
$J^2 = -1$, we have the ordinary complex numbers $\mathbb{C}$, and
if $J^2 = +1$, we have the split complex numbers
$\mathbb{C}^{\prime}$; for $\mathbb{C}$ and also for
$\mathbb{C}^{\prime}$ we have the involutory automorphism  $z =
x+Jy \mapsto \overline{z}= x-Jy$, with $x, y$ reals. Then $Q(z)
:=\overline{z} z\in\mathbb{R}$ is a quadratic form, and as
$Q(zz^{\prime})=Q(z)Q(z^{\prime})$, we have two possible
composition algebras, for the complex numbers $\mathbb{C}$ and for
the split complex $\mathbb{C}^{\prime}$. For $\mathbb{C}$, we have
a field (=commutative division algebra), because $z^{-1}
:=\overline{z} /Q(z),\; z\neq 0$, makes sense, whereas in
$\mathbb{C}^{\prime}$, the split case, $Q(z)=0$ implies that $z$
lies in any place in the "light cone" $x = \pm y$. So both the
complex and the split complex make up a composition algebra, but
only the complex $\mathbb{C}$ are also a division algebra and it
is a field, as it is commutative. For both $\mathbb{C}$ and
$\mathbb{C}^{\prime}$ we have the two-squares identity, which are
just the expression of the composition property
$Q(zz^{\prime})=Q(z)Q(z^{\prime})$: for $z = x + Jy, z^{\prime}=
x^{\prime}+Jy^{\prime}$, $J^2 =\pm 1$; known from antiquity, they
are
\begin{eqnarray}
\left(x^{2}+y^{2}\right)\left(x^{\prime 2}+y^{\prime
2}\right)=\left(x x^{\prime}-y y^{\prime}\right)^2+ \left(x
y^{\prime}+y x^{\prime}\right)^2\quad {\rm from\; } \mathbb{C}\;
(J^2=-1)\label{Eq21}\\
\left(x^{2}-y^{2}\right)\left(x^{\prime 2}-y^{\prime
2}\right)=\left(x x^{\prime}+y y^{\prime}\right)^2- \left(x
y^{\prime}+y x^{\prime}\right)^2\quad {\rm from\; }
\mathbb{C}^{\prime}\; (J^2=+1)\label{Eq22}
\end{eqnarray}

If $\mathbb{C}^{*} = \mathbb{C}-\left\{0\right\}$ is the
multiplicative group of the field $\mathbb{C}$, we have
topologically $\mathbb{C}^{*}\simeq
\mathbb{S}^{1}\times\mathbb{R}^{+}$ (polar decomposition), where
$\mathbb{S}^{1}$ are just the $Q(z)=1$ complex numbers and
$\mathbb{R}^{+}$ the positive reals. The automorphisms of both
$\mathbb{C}$ and $\mathbb{C}^{\prime}$ as $\mathbb{R}$-algebras
are just complex conjugation, Aut($\mathbb{C}$) =
Aut($\mathbb{C}^{\prime}$) = $\mathbb{Z}_2 = O(1)\simeq
\mathbb{S}^{0}$ ($O(n)$ is the $n$-dim real orthogonal group,
$n\geq 1$, and $\mathbb{S}^{n}$ the $n$-sphere, $n\geq 0$); the
fix point set is the real line (Recall the complex {\it as a
field} has other (discontinuous) automorphisms).

The story is analogous in the next step, the (split) quaternions
(Hamilton, 1843). One has now two units, $i$ and $j$, both of
square $\pm 1$; then $k:= ij$ has to be another unit, so one has
the vector space $\mathbb{R}^4$. For the division algebra case,
$\mathbb{H}$, $i^2=j^2=k^2= -1$, the three $e$: $i,j$ and $k$ are
anti-involutory ($e^4=1$) and anticommuting, and then, if $q$ is a
generic quaternion, $q = u + \boldsymbol{\sigma\cdot x}$ = real
plus imaginary, where $\boldsymbol{\sigma} = \left\{i,j,k\right\}$
and $\textbf{x}$ is a real $3$-vector, the conjugation
$q\mapsto\overline{q}$ is an (anti-)automorphism, and again
$Q(q):= \overline{q}q\in\mathbb{R}$ is a quadratic form, definite,
$Q(q) = u^2 + \bf{x}\cdot \bf{x}$: sum of real squares, satisfying
$Q(qq^{\prime})=Q(q)Q(q^{\prime})$. The inverse is $q^{-1} =
\overline{q}/Q(q),\; q\neq 0$, and one has the division algebra of
the quaternions $\mathbb{H}$. Notice the conventional vector
product in $3$-dim is $\textbf{x}\wedge \textbf{y} = {\rm
Im}(\textbf{xy}) =\left[\textbf{x}, \textbf{y}\right]/2$, where
$\textbf{x}, \textbf{y}$ are imaginary quaternions, and the scalar
product verifies $\textbf{x}\cdot \textbf{y} = - {\rm
Re}(\textbf{xy})$. The pair $q = (u, \textbf{x})$ was already
called by Hamilton the scalar $(u)$ and the vector $(\textbf{x})$
part of the quaternion; the "noncommutative field" of the
quaternions is named at times a skew-field.

To get the split quaternions $\mathbb{H}^{\prime}$  one takes one
of the units involutory, e.g. $i^2 = +1$. Then everything works
similarly as in the split complex case, with now $j^2 = -1$ and
$k^2 = (ij)^2 = +1$, and one still has a composition algebra; but
the quadratic form is of $(++--)$ signature:
$Q(q)=u^2-x^2+y^2-z^2$; (if the three units have square $+1$, they
do not anticommute, and there is \textit{no} composition
algebra!).

It is interesting to look at (anti)automorphisms of these algebras
$\mathbb{H}$ and $\mathbb{H}^{\prime}$.  For the (true)
quaternions $\mathbb{H}$, it is easy to see that any rotation in
the $3$-space of unit imaginary $q$'s is an automorphism, and
viceversa, so Aut($\mathbb{H}$)$=SO(3)$: any frame $\epsilon(i, j,
k=ij)$ is rotated in another one $\epsilon^{\prime}$ with the same
orientation. Conjugation is an antiautomorphism, i.e.
$\overline{(qq^{\prime})} = \overline{q^{\prime}}\overline{q}$ ,
so the group of autos and antiautos  AntiAut($\mathbb{H}$) is
$O(3)$. There is also the group $\mathbb{Z}_3 = A_3$ permuting
cyclically the imaginary units $i, j$ and $k$, as it is an
oriented frame. The $SO(3)$ automorphism group leaves the
imaginary volume form (a $3$-form) invariant; the quadratic form
$Q(q)$ is invariant under $O(4)$, of course.

For the split quaternions $\mathbb{H}^{\prime}$ the automorphism
group is the connected part of the orthogonal group for the
$(++-)$-metric, $SO_{0}(2, 1)$, and there is no cyclic symmetry.
In the full split quaternions $\mathbb{H}^{\prime}$ the metric is
of signature $(2, 2)$, so the isometry group of $Q(q)$ is now
$O(2, 2)$. Unit quaternions make up $\mathbb{S}^3=SU(2) = Sq(1)=$
Spin($3$) (the double and universal cover of $SO(3)$), and can be
used efficiently to describe rotations in $3$- and $4$-space
\cite{Con}. We also have the "polar" decomposition $\mathbb{H}^{*}
= \mathbb{H}-\left\{0\right\}\simeq \mathbb{S}^3 \times
\mathbb{R}^{+}$, where $\mathbb{S}^3\simeq SU(2)$. We name $Sq(n)$
the "unitary" $n\times n$ group, with quaternion entries.

It is remarkable that at this level, for the composition character
only anticommutativity  $\left\{i, j\right\}=0$ plays a role,
whereas for the division property one needs also the
anti-involutory condition, that is, $J^2 = -1$ for the three units
$i, j, k$.

We omit the corresponding two four-squares identities (Euler) for
$\mathbb{H}$ and $\mathbb{H}^{\prime}$, analogous to (\ref{Eq21})
and (\ref{Eq22}), for brevity.

The application of complex numbers in physics needs no apology; we
only mention here that the very first formulations of Quantum
Mechanics incorporate the $i$, both in Heisenberg´s matrix
mechanics (June, 1925) as in the more widespread Schr\"odinger
wave equation (January, 1926).

Quaternions are used in the $3$-dim vector calculus, as stated
above, although generally without mentioning the origin in the
quaternion algebra. Rotations in Quantum Mechanics again use the
covering group, $Sq(1) = $Spin($3$), but the usual notation, as
$SU(2)$, makes it easy to forget the true nature is really the
unit quaternions. But quaternion product as a way to quickly
compose $3$-dim rotations are used in many devices, even in some
electronic games...

\section{Octonions and split octonions}

If we now have three algebraically independent units $i, j, k$,
one has to go to $\mathbb{R}^8$, with units $1; i, j, k; ij, jk,
ki; (ij)k$. For the (true) octonions $\mathbb{O}$ the seven units
anticommute and square to $-1$.  Define $o := u +
\boldsymbol{\rho}\cdot \xi$, (8=1+7), where $u$ is real, $\xi$ a
real $7$-vector, and $\boldsymbol{\rho}$ is a short for the seven
imaginary units; define routinely the conjugate as $\overline{o} =
u - \boldsymbol{\rho}\cdot\xi$, so $Q(o):=\overline{o} o = u^2
+\xi^2\geq 0$, and inverse as $o^{-1} = \overline{o}/Q(o),\; o\neq
0$. One sees at once that in order $\overline{o}o$ to be real you
need {\it alternativity}: $i(jk) = -(ij)k$ \textit{en lieu} of
associativity: this is the peculiar property of the octonions (and
what made Hamilton to reject them). Just here composition
character implies alternativity (and a hint of Hurwitz theorem:
you cannot proceed beyond dim 8 with composition algebras),
whereas the division condition would imply again anti-involutory
character of all seven imaginary units. Then $o ={\rm Re\;} o
+{\rm Im\;} o$ (8=1+7), with norm $N(o)=\left|o\right|=
+\sqrt{\overline{o} o}$; in this division algebra case the $Q$ is
definite. Another way to obtain alternativity is to insist that
two of the three units $(i, j, k)$ generate a quaternion algebra.
The precise definition of alternativity is this: an algebra $A$ is
{\it alternative}, if the alternator $\left[a, b, c\right]:= (ab)c
- a(bc)$ is fully antisymmetric in the three arguments: any
associative algebra is alternative, of course, as $\left[ a, b,
c\right]= A - A = 0$. The polar decomposition is again
$\mathbb{O}^{*}\simeq \mathbb{S}^7\times \mathbb{R}^{+}$, on the
understanding that $\mathbb{S}^7$ is {\it not} a group (it has a
structure called a {\it loop}). We recall that the four spheres
$\mathbb{S}^0, \mathbb{S}^1, \mathbb{S}^3$ and $\mathbb{S}^7$ are
the only parallelizable ones, and clearly described the $Q(x)=1$
numbers for $x$ in $\mathbb{R}, \mathbb{C}, \mathbb{H}$ and
$\mathbb{O}$ respectively.

For the split octonions $\mathbb{O}^{\prime}$ we can take e.g.
$(e_1=i)^2=+1$, and then $(e_1e_2)=e_4$, $(e_3e_1) =e_6$ and $e_7$
are also involutive, so the $7$-dim metric has signature
$(+--+-++)$, and $Q = \overline{o} o$ has $(4, 4)$ signature; of
course, one still needs alternativity to guarantee $Q(oo^{\prime})
= Q(o)Q(o^{\prime})$.

The situation is thus nearly identical for the three division
algebras $\mathbb{C}, \mathbb{H}$ and $\mathbb{O}$: the new units
are anti-involutory, and have to anticommute in this, the division
algebra case; $\mathbb{C}$ is commutative and associative,
$\mathbb{H}$ is associative but not commutative, and $\mathbb{O}$
is neither. Anticommutativity restricts non-commutativity for
quaternions, and alternativity substitutes non associativity for
octonions; as there are no more properties to sacrify, there are
neither composition nor divison algebras in higher dimensions,
although there are some rings, called e.g. {\it sedenions} (dim
$16$) for the next case, of four algebraically independent units:
if one proceeds to these sedenions, with four units $i, j, k, l$,
and ($1+4+6+4+1$)$=16$ dimensions, the possible algebras are
neither composition nor division (Hurwitz, 1896; Zorn, 1930). The
process, which is a variant of the Cayley-Dickson method
\cite{Baez}, continues...

A nice way to represent the octonions is the Fano plane
$\mathbb{F}_2\mathbb{P}^2$ (see e.g. \cite{Baez}), the projective
plane over the (Galois) field of two elements $\mathbb{F}_2$: it
draws the $7 = p^2 + p +1$ (for $p=2$) imaginary units in a
triangle (three vertices, three mid-lines and the center). Recall
the order of the (projective) symmetry group $PGL_3(2) = PSL_2(7)$
is $168 = (2^3-1)(2^3-2)(2^3-4) = (7^2-1)(7^2-7)/2$, and it is the
second smallest non-Abelian simple group, the first one being $A_5
= PSL_2(5)$, of order $60 =5!/2$.

What about octonion automorphisms? The possible group
Aut($\mathbb{O}$) has to lie inside $SO(7)$, with dim $21$,
because there is norm and orientation to preserve; let us just
call Aut($\mathbb{O}$)$ :=G_2$; it has to be smaller than $SO(7)$,
because not any $7$-frame goes into any other: both Baez
\cite{Baez} and Rosenfeld \cite{Rosen} make easy the case for
$G_2$ to have dimension $14$; taking $i= e_1$ in the $6$-dim
sphere of unit imaginary octonions, $j= e_2$ has to lie in the
orthogonal equator ($\mathbb{S}^5$), and $k= e_3$ in the $3$-dim
space orthogonal to ($i, k$ and $ij$): so dim Aut($\mathbb{O}$)$=
6+5+3=14$ (this shorter argument is also in (\cite{Con}, p. 76)).
Then, as $G_2$ acts transitively in the $6$-sphere of unit
imaginary octonions, the isotropy group $K\subset G_2$ has
dimension $14-6=8$: it turns out to be the $SU(3)$ group with the
real irreducible $6$-dim representation (becoming  $\overline{3}
+3$ over the complex): $SU(3)$ leaves the $\mathbb{S}^5$ diameter
of the previous $6$-sphere fixed, embeddable in a $\mathbb{R}^6$
space; we have then $G_2/SU(3)\simeq \mathbb{S}^6$ as an
homogeneous space (it turns out {\it not} to be a symmetric
space), and also $SU(3)/SU(2) = \mathbb{S}^5$, as the exact
sequence is $SU(2)\rightarrow SU(3)\rightarrow \mathbb{S}^5\subset
\mathbb{R}^6$. See diagrams at the end of the paper, in Sect. 6.

What about a discrete group, in this octonionic case, which would
play the role of the $\mathbb{Z}_3 = A_3$ in the quaternion case?
The seven imaginary units can be cyclically permuted, which gives
a $\mathbb{Z}_7$ group, but this is not all: the Fano projective
plane is also the projective line over $\mathbb{F}_7$; one has
also triangular $2\pi /3$ rotations as symmetries, and they
combine with $\mathbb{Z}_7$ to make up a non-Abelian group of
order $21$. The natural group acting on the Fano plane is
$PGL_2(7)$, of order $336$, while $\left|PSL_2(7)\right|= 168$.
Our discrete group is thus
\begin{equation}
\mathbb{Z}_7 \ltimes \mathbb{Z}_3,\label{Eq31}
\end{equation}
where $\mathbb{Z}_3\subset \mathbb{Z}_6 =$ Aut($\mathbb{Z}_7$)
determines the semidirect product. This corresponds to the $A_3$
case for the quaternions, but here $\mathbb{Z}_7\ltimes
\mathbb{Z}_3\subset A_7$ ; incidentally, this $21$ order
non-Abelian group is the only other, with the direct sum abelian
group $\mathbb{Z}_3\oplus \mathbb{Z}_7$, of this order.

Finally unit imaginary octonions form $\mathbb{S}^6$, which admits
a quasi-complex structure (Borel-Serre) due to imaginary octonion
multiplication: among the even spheres only $\mathbb{S}^2$ and
$\mathbb{S}^6$ admit a (quasi-)complex structure, truly complex
for $\mathbb{S}^2$.

Incidentally, it was Richard Feynman who first established a kind
of $7$-dimensional vector product, $(\boldsymbol{\xi}\wedge
\boldsymbol{\eta})=\left[\boldsymbol{\xi},\boldsymbol{\eta}\right]/2
= {\rm Im\;}(\boldsymbol{\xi} \boldsymbol{\eta} )$, for
$\boldsymbol{\xi},\boldsymbol{\eta}$ imaginary octonions,
generalizing the Hamilton-Gibbs vector calculus in three
dimensions \cite{Feyn}.

The automorphism group of the split octonions
$\mathbb{O}^{\prime}$ is the noncompact real form of $G_2$, which
lives inside $SO_{0}(4, 3)$. The split octonion quadratic form
$Q^{\prime}$ admits the $O(4, 4)$ group as isometry, of course.

Modern high-energy physics uses many groups associated with the
octonions, as $G_2$ (holonomy of compactifying from $11$-dim
space), $E_8$ as gauge group of the M-Theory still in $11$-dim,
not to speak of $E_8^2$, used in string theory, or $E_{6}$, which
appears in Grand Unification Theories (GUTs).

\section{Some classes of tensors}

We start now a seemingly totally independent development. Consider
$T^p_q$ (=tensors on a vector space, say $V$ over a field
$\mathbb{F}$); take a particular one, $t\in T^p_q$; imagine the
general linear group, with $g\in GL(V)$ acting on it in the
natural way, write $g\cdot t=t^{\prime}$, and try to classify the
tensors by equivalence classes (orbits) under the $GL$ action: as
$\dim GL(V) = n^2$, the simple vectors $x$ (as $T^1_0$ tensors)
are classified: the zero vector and the rest; the little group of
the later orbit is the affine group $A_{n-1}$. Rank two covariant
tensors $T^0_2$ ($\Longleftrightarrow$ bilinear forms) split first
into symmetric and antisymmetric parts, and there is also regular
character (or not): any bilinear form $b\in T^0_2$ generates a
linear map from $V$ to the dual space $V^{*}$, $b^{\prime}:
V\rightarrow V^{*}$, and as they have the same dimension, the form
$b$ is called regular if the map $b^{\prime}$ is an isomorphism
(for good reasons, Hitchin \cite{Hit2} speaks of {\it stable}
forms instead of regular). The isotropy groups for the regular
case give rise to the orthogonal and symplectic groups in the
known way. We recall:

If $Q$ is a regular quadratic form, the isotropy group $\left\{
g,\; g\cdot Q = Q\right\}$ defines an orthogonal group $O(n)$; if
$\omega$ is a regular $2$-form, the group is the symplectic one,
$Sp(n)$ (acting on even dimension $2n$ space); if $\tau$  is a
volume form (or $n$-form $\neq 0$), the isotropy group is the
unimodular group, $SL(n)$: so over any field, we have the three
classical series of matrix groups of Cartan: $B$-$D$, $C$ and $A$;
for some fields, e.g. over the reals, there is a further
distinction by the Sylvester signature.

The natural question now arises: Is there any other orbits
possible, tensors under $GL(V)$, besides the obvious
$Q,\omega,\tau$ and with isotropy groups? If so, which (new)
groups arise? For endomorphisms End($V$)$\simeq T^1_1$ the orbits
under $GL(n)$ are the so-called "elementary divisors", which
classify matrices; their little groups are easily identified, and
are not very interesting for our purposes. The next possible case
are $p$-forms \cite{Hit}, with dimension $\dim \bigwedge
T^0_p(\mathbb{F}^{n})= {n \choose p}$, but as ${9 \choose 3} = 84
> 9^2=81$, we have potentially four cases only:

\begin{itemize}

\item  $3$-forms on $\mathbb{F}^7 \quad (7^2 = 49 > {7\choose
3}=35)$.

\item $3$-forms on $\mathbb{F}^8\quad (8^2=64 > {8 \choose
3}=56)$.

\item (Self)-$3$-forms in $\mathbb{F}^6\quad (6^2=36 > {6 \choose
3}=20 > \frac{1}{2}{6 \choose 3} =10)$.

\item Self-$4$-forms in $\mathbb{F}^8 (8^2=64 > \frac{1}{2}{8
\choose 4 }=35$.
\end{itemize}

As we have  $\frac{1}{2}{2n\choose n}= {2n-1 \choose
n-1}={2n-1\choose n}$, we have dim (self-$3$-forms in
$\mathbb{F}^6$)$ = 10 =$ dim ($2$-forms in $\mathbb{F}^5$), but
the $3$-forms themselves are very interesting, the isotropy group
being $SL_3(\mathbb{C})^2$ for $\mathbb{F}=\mathbb{C}$ ($6^2 -
2\cdot 10 = 16$) and $SL_3(\mathbb{C})$ for
$\mathbb{F}=\mathbb{R}$, \cite{Hit}. For $\mathbb{F}=\mathbb{C}$
the really interesting case is the generic $3$-forms in
$\mathbb{C}^7$. So for a generic $3$-form $\phi$, we have dim
Aut($\phi$)$= 49-35=14$, the same as the dimension of complex
$G_2$! In fact, Bryant \cite{Bry} bases his study of $G_2$ as
invariance group of $\phi$. Engel and Reichel (see above)
determined that there are two regular (stable) real forms, with
isotropy the compact and noncompact forms of real $G_2$.

Is there any sensible isotropy group for a generic $3$-form in
$\mathbb{R}^8$? It will be of dimension $8^2 -{8 \choose 3} = 64 -
56 = 8$, and in the compact case the candidate would be
$SU(3)/\mathbb{Z}_3 = \mathbb{PU}(3)$, which has  a nice $8$-dim
real representation, in fact the adjoint of $SU(3)$; but we do not
consider this case anymore (see again \cite{Hit,Hit2}), except the
trivial remark that then the $SU(3)$ group would appear a second
time in relation to $\mathbb{R}^8$ (and the octonions), with
possible applications in physics: both flavour and colour physics
use the group $SU(3)$ consistently.

The (anti-)self-dual $4$-forms are classes under
$SL_8(\mathbb{R})$ and have also dimension $35$; they are called
Cayley forms. They are not generic; the pertinent group happens to
be Spin($7$), which has also a single real $8$-dim representation;
a nice discussion is in (\cite{Joy}, p. 255), as Spin($7$),
together with $G_2$, are the two exceptional holonomy groups. In a
precise way, which we do not elaborate, the $3$-form in
$\mathbb{F}^7$ comes really from the self and antiselfdual forms
in $\mathbb{F}^8$.

The proof that there are no more left over cases is just
numerical: $n^2 = \dim GL_n(\mathbb{F})$ is less than any other
generic tensor; the reader can be convinced by himself that
generic tensors of any other rank would be equivalent to the above
ones or have no room for isotropy groups.

\section{The $3$-form in dimension $7$ and the octonions}

Starting with a generic $3$-form $\phi$  in $\mathbb{F}^7$, if
there is such a thing, the isotropy group, of dimension $14$, will
be a certain group $G:= G(3; 7)$. One suspects, of course, that
for $\mathbb{F}=\mathbb{R}$, $G(3; 7) =$ Aut($\mathbb{O}$); how
does one prove this? We shall only indicate the idea of the proof.
Start by the $3$-form in $\mathbb{F}^7$ (for $\mathbb{F} =
\mathbb{R}, \mathbb{C}$) and try to recover the octonions, with
the group $G_2$ playing the dual role: it is the isotropy group of
a "generic" $3$-form in $\mathbb{F}^7$ and at the same time the
automorphism group of the (constructed) octonion algebra. The idea
of the proof goes along the following steps:

\begin{enumerate}

\item There is also a sense of regularity in the $3$-form: in
fact, a such form $\phi$, generates an special bilinear symmetric
form $\beta$ (found already by Engel, see \cite{Agric,Hit}): for
$\bf{x}, \bf{y}$ vectors in $\mathbb{F}^7$, if $\phi(\bf{x})$ is
the $2$-form contraction, $\phi(\bf{x}) = \bf{x}\lrcorner \phi$,
\begin{equation}
\beta(\bf{x},\bf{y}) := \phi(\bf{x})\wedge\phi(\bf{y})\wedge\phi,
\quad {\rm (i.e., a\; 7-form)}
\end{equation}

\item Call $\phi$  regular if $\beta(\phi)$ is regular
(non-degenerate). The isotropy group of this special
$7$-dimensional metric is $SO(7, \mathbb{F})$, and one recovers
the natural inclusion $G(3;7)\subset SO(7, \mathbb{F})$. Engel
proved that, in the three cases ($\mathbb{F}= \mathbb{C}$ and the
two $\mathbb{F}=\mathbb{R}$), the $\beta$ form is non-degenerate
for a generic (=lying in an open set) $3$-form $\phi$.

\item The $SO$ group lies inside $SL(\mathbb{F})$, hence there is
a volume element; therefore, there is the Hodge duality operator
$^{*}$, and there is also an invariant 4-form $\psi=^{*}\phi$, as
$\phi\wedge ^{*}\phi=\tau$ is the volume element. Recall dim
$3$-forms = dim $4$-forms for $n =7$.

\item With the regular metric  $\beta$ and the $3$-form $\phi$,
one gets an algebra! This is because an algebra is a (particular)
$T^{1}_{2}$ tensor, as $xy=z$ means precisely this;  the metric
being regular, $\exists\beta^{-1}$, and one flips an index,
passing from $\phi$ a $\wedge T^{0}_{3}$ tensor to a $T^{1}_{2}$
one; write $\beta^{1}(\phi)=\gamma$. Hence, in $\mathbb{F}^7$ one
has an algebra, with $\gamma(\bf{x},\bf{y}) = \bf{z}$. It is
antisymmetric, $\gamma(e_1,e_2)= e_1e_2 = - e_2e_1$ (for a basis
$e_1,...,e_7$), because so is $\phi$  in two indices, and it is
alternative, because $\phi$ is fully antisymmetric! See 8) below.

\item One has reproduced the "vector" product of imaginary
octonions!, and so you can reconstruct the Fano plane for
imaginary octonionic multiplication.

\item Adding now the unit 1, so that $e_i^2$ is not zero but $-1$
(or $+1$, see below), we have reconstructed the octonion
composition and division algebras in the case
$\mathbb{F}=\mathbb{R}$! And of course, now it is clear that the
$G_2$ group has two faces: it is either the isotropy group of a
$3$-form, or automatically the Aut group of the algebra: this is
the looked-for {\it Two Faces of} $G_2$. One shows also that $G_2$
is compact for the octonion division case ($\mathbb{O}$), and a
noncompact form, lying in $SO_{0}(4, 3)$, for the split case,
where some of the units square to $+1$. Engel and Reichel proved
\cite{Agric} that there is a unique generic class of generic
$3$-forms for $\mathbb{F}=\mathbb{C}$, and two for
$\mathbb{F}=\mathbb{R}$.

\item Comparing this with the (true) quaternions $\mathbb{H}$,
there the $3$-form   is the volume form, with $SL_3(\mathbb{R})$
as isotropy group; but the metric is put by hand, so the group
becomes $SL_3\cap O(3) = SO(3)$; and indeed
Aut($\mathbb{H}$)$=SO(3)$.

\item For the concrete form of the $3$-form one mimics the
octonion product: if $\left\{e_i\right\}$ is a basis in, say,
$\mathbb{R}^7$, with  $\omega_i$ the dual basis,
$\omega_i(e_j)=\delta_{ij}$, the generic $3$- form can be defined
\cite{Bry,Joy} as
\begin{equation}
\phi = (124) +  ( 157) +  (163) + (235) +  (276) + (374) +
(465),\label{Eq52}
\end{equation}
where $(124)$ means $\omega_1\wedge\omega_2\wedge\omega_4$ etc.
Passing through the metric $\beta$ and to the algebra $\gamma$,
one should reproduce the octonion product of Sect. 3, to wit
\begin{equation}
e_1e_2 = e_4,\; e_2e_3 = e_5, ...\label{Eq53}
\end{equation}

\item The argument above is oversimplified; for example, in the
same way that a bilinear form over the reals might have a
signature (Sylvester), it turns out that there are two cases of
generic $3$-forms in $\mathbb{R}^7$ (Engel, Reichel). So one goes
to the octonions and to the split octonions, as the isotropy group
of these forms is compact and noncompact respectively.

\item Notice the difference between the two cases (corresponding
to the true and split octonions) comes from adding the units,
$e_i^2$, and does not depend much on the $7$-dim structure (see 6
above).

\end{enumerate}

\section{Spin groups}

We conclude by establishing the following relations between the
several groups appearing in our study, inspired in (\cite{Joy}, p.
256); we first recall the coincidences among the small dimension
spin groups (see e.g. \cite{Harv}):
\begin{eqnarray}
{\rm Spin\;}\left(1\right)=O(1);\; {\rm Spin\;}\left(
2\right)=SO(2)=U(1);\; {\rm Spin\;}\left(
3\right)=SU(2)=Sq(1);\nonumber\\
{\rm Spin\;}\left( 4\right)={\rm
Spin\;}\left(3\right)^2;\; {\rm Spin\;}\left( 5\right)=Sq(2);\;
{\rm Spin\;}\left( 6\right)=SU(4).
\end{eqnarray}

We consider now the five groups with an irreducible representation
of real dimension $8$, namely $SO(8)$, Spin($8$), $SU(4)$, $Sq(2)$
and Spin($7$); they are clearly related to $\mathbb{R},
\mathbb{C}, \mathbb{H}$ and $\mathbb{O}$; they also act
transitively on the $7$-sphere $\mathbb{S}^7$. We shall consider
three diagrams to illustrate these relations. Figure 1. shows the
relation of the quaternions and the complex numbers, Figure 2.
that of the octonions and the complex numbers, while considering
the reals and the octonions yields the following diagram of Figure
3.:

\begin{figure}[h]\footnotesize
\caption{$\mathbb{H}$ and $\mathbb{C}$}
\begin{diagram}
{\rm Spin\;}\left(  3\right) = SU\left(  2\right) = Sq\left(
1\right)& \longrightarrow & Sq\left(  2\right) = {\rm Spin\;}
\left( 5\right)& \longrightarrow & \mathbb{S}^{7}\subset \mathbb{R}^{8}= \mathbb{H}^{2}\\
  \dTo & & \dTo & & \\
SU(3)& \rTo & SU(4)={\rm Spin\;}(6) & \rTo & \mathbb{S}^{7}\subset \mathbb{R}^{8}= \mathbb{C}^{4}\\
\dTo & & \dTo & & \\
\mathbb{S}^{5} & \rEq & \mathbb{S}^{5} & & \\
\end{diagram}
\end{figure}

\begin{figure}[h]\footnotesize
\caption{$\mathbb{O}$ and $\mathbb{C}$}
\begin{diagram}
SU\left(  3\right)& \rTo & SU(4)= {\rm Spin\;}\left( 6\right)&
\rTo
& \mathbb{S}^{7}\subset \mathbb{R}^{8}= \mathbb{C}^{4}\\
  \dTo & & \dTo & & \\
G_{2}& \rTo & {\rm Spin\;}(7) & \rTo & \mathbb{S}^{7}\subset \mathbb{R}^{8}= \mathbb{H}^{2}\\
\dTo & & \dTo & & \\
\mathbb{S}^{6} & \rEq & \mathbb{S}^{6} & & \\
\end{diagram}
\end{figure}

\begin{figure}[h]\footnotesize
\caption{$\mathbb{R}$ and $\mathbb{O}$}
\begin{diagram}
G_2 & \rTo & {\rm Spin\;}\left( 7\right)& \rTo
& \mathbb{S}^{7}\\
  \dTo & & \dTo & & \dEq \\
SO(7) & \rTo & SO(8) &\rTo & \mathbb{S}^{7}\\
\dTo & & \dTo & & \\
\mathbb{RP}^{7} & \rEq & \mathbb{RP}^{7} & & \\
\end{diagram}
\end{figure}

Only the first two columns require an explanation: as obviously
Spin($8$)/Spin($7$)$ = \mathbb{S}^7$, and Spin($8$) covers $SO(8)$
twice, we have $SO(8)/$Spin($7$)$=$ the real projective space
$\mathbb{RP}^7\simeq \mathbb{S}^7/\mathbb{Z}_2$.

The relation between spin groups and division algebras, in
particular the octonion algebras, is noteworthy; we elaborate this
in \cite{Boy}. Lastly, a remark on Spin($8$), which exhibits a
wonderful triality: the extension of $O(8)$ by the three $8$-dim
representations (permuted by triality) gives rise to $F_4$, the
second exceptional group, (${8\choose 2} + 3\cdot 8= 52 = \dim
F_4$), whereas the quotient of $O(8)$ by the same triality
automorphism group generates $G_2$ (again!): the triple bond in
the Dynkin diagram for $G_2$ (see Sect. 1) comes really from
triality in $O(8)\simeq D_4$! \cite{Ada}.

The center of Spin($8$), as any Spin($4n$) group, is $V =
\mathbb{Z}_2\oplus \mathbb{Z}_2$: and we know that Aut($V$) is the
symmetric group $S_3$; in all cases except Spin($8$) (and PO($8$)
= Spin($8$)$/V$) the outer automorphisms of the group shrinks to
$\mathbb{Z}_2$ (which e.g. permutes the two chiral
representations), but in Spin($8$) the $S_3$ group of autos of the
center lifts to a $S_3$ group of autos (triality!) of the full
group: the deep reason of this is the "loop" multiplicative
character of the seven sphere of unit octonions, which appears
twice in Spin($8$): the sphere homology product of Spin($8$) is
$\mathbb{S}^3\odot \mathbb{S}^7 \odot \mathbb{S}^7 \odot
\mathbb{S}^{11}$.

This triality is at the base of some supersymmetric field theories
in physics, in particular $\mathcal{N}=1$ Susy Yang-Mills in $10 =
(1, 1) + (8, 0)$ dimensions, as well as the string theory IIA,
with $\mathcal{N} = 2$ Supersymmetries, involving the vector and
the two spinor representations for the light cone group $O(8)$ as
dictated by triality; see \cite{Har}. So in conclusion, as regards
to physics, both the groups $SU(3)\subset G_2$ and triality (as
enhancement of duality), which is connected with octonions, seem
to be related, if not unavoidable, in modern physical theories.

Even the group $F_4$ does appear in an unexpected place in
physics: namely, in M-Theory or rather in the low-energy particle
content. P. Ramond as shown \cite{Ram} that the particles $h$
(graviton, dim $44$), $\Psi$ (gravitino, dim $128$) and $C$
($3$-form, dim $84$) are related to the Moufang octonionic plane,
$\mathbb{OP}^2$, which as symmetric space is $F_4/$Spin($9$): the
three represenations are induced from the Id representation of
$F_4$, where the triplet is related to the Euler number, as
$\chi(\mathbb{OP}^2) =3$. For further developments and extensions
to F-Theory, see \cite{Boya}.

\section*{Acknowledgements}

The first author (LJB) acknowledges partial financial support by
the CICYT grant FPA2006-0235. The second author (RCS) acknowledges
support by the research grant GR58/4120818-920920 of the UCM-BSCH.

\end{document}